
\input amsppt.sty
\magnification=\magstep1
\hsize = 6.5 truein
\vsize = 9 truein

\NoRunningHeads
\NoBlackBoxes
\TagsAsMath
\TagsOnRight
\catcode`\@=11
\redefine\logo@{}
\catcode`\@=13

\def\label#1{\par%
        \hangafter 1%
        \hangindent .66 in%
        \noindent%
        \hbox to .66 in{#1\hfill}%
        \ignorespaces%
        }

\newskip\sectionskipamount
\sectionskipamount = 24pt plus 8pt minus 8pt
\def\sectionskip{\vskip\sectionskipamount}
\define\sectionbreak{%
        \par  \ifdim\lastskip<\sectionskipamount
        \removelastskip  \penalty-2000  \sectionskip  \fi}
\define\section#1{%
        \sectionbreak   
        \subheading{#1}%
        \bigskip
        }


\define\op#1{\operatorname{\fam=0\tenrm{#1}}} 

        \define         \x              {\times}
        \let            \< = \langle
        \let            \> = \rangle
        \define         \a              {\alpha}
        \redefine       \b              {\beta}
        \redefine       \d              {\delta}
        \redefine       \D              {\Delta}
        \define         \e              {\varepsilon}
        \define         \g              {\gamma}
        \define         \G              {\Gamma}
        \redefine       \l              {\lambda}
        \redefine       \L              {\Lambda}
        \define         \n              {\nabla}
        \redefine       \var            {\varphi}
        \define         \s              {\sigma}
        \redefine       \Sig            {\Sigma}
        \redefine       \t              {\tau}
        \define         \th             {\theta}
        \redefine       \O              {\Omega}
        \redefine       \o              {\omega}
        \define         \z              {\zeta}
        \redefine       \i              {\infty}
        \define         \p              {\partial}

\topmatter
\title Quasitriangularity of Quantum Groups at Roots of 1
\endtitle
\author N. Reshetikhin\footnote{This work was supported by an Alfred
  P. Sloan fellowship and by National Science Foundation grant
  DMS--9296120.\newline}
\endauthor
\abstract
   An important property of a Hopf algebra is its quasitriangularity
   and it is useful for various applications. This property is
   investigated for quantum groups $sl_2$ at roots of 1.
   It is shown that different forms of the quantum group $sl_2$ at
   roots of 1 are either quasitriangular or have similar structure
   which will be called autoquasitriangularity.In the most interesting cases
   this property means that "braiding automorphism" is a combination of
   some Poisson transformation and an adoint transformation with certain
   element of the tensor square of the algebra.

\endabstract
\address\hskip-\parindent
        Department of Mathematics \newline University of California \newline
        Berkeley, California 94720
\endaddress
\endtopmatter
\baselineskip=14 pt
\document

Algebras which here will be called quantum $sl_2$
are the simplest
examples of quantum groups which have practically all the remarkable
properties of this class of Hopf algebras.  One of the most important
properties of quantum groups is quasitriangularity.  Recall the
definition from [Dr].

\bigskip
{\bf Definition 1.} A Hopf algebra $A$ is called quasitriangular if
there exists $R\in A\otimes A$ (or an element
>from the appropriate completion of $A\otimes A$) such that
$$
\Delta'(a) = R\Delta (a)R^{-1}    \tag 1
$$
$$
(\Delta\otimes{\op{id}})(R) = R_{13}R_{23}    \tag 2
$$
$$
({\op{id}}\otimes\Delta)(R) = R_{13}R_{12}    \tag 3
$$
Here $\Delta'(a)=\sigma\circ\Delta(a)$ where
$\sigma: A^{\otimes 2}\to A^{\otimes 2}$,
\ $a\otimes b\mapsto b\otimes a$ and
$R_{12},R_{13},R_{23}\in A^{\otimes 3}$ (or to the appropriate
completion of it), $R_{12}=R\otimes 1$, \ $R_{23}=1\otimes R$, \
$R_{13}=(\sigma\otimes{\op{id}})(R_{23})=({\op{id}}\otimes\sigma)(R_{12})$.

\bigskip
A remarkable corollary of this definition is that $R$ satisfies the
Yang-Baxter equation in $A^{\otimes 3}$:
$$
R_{12}R_{13}R_{23} = R_{23}R_{13}R_{12}
$$

It is known [Dr] that quantum universal enveloping algebra
$U_h \frak g$ are quasitriangular over $\Bbb C[[h]]$ for any
Kac-Moody algebra $\frak g$. It is also known that the corresponding
algebraic quantum universal algebra $U_q(\frak g)$ [J] are not
quasitriangular over $\Bbb C[q,q^{-1}]$.
This fact is essential when
$q$ is specialized at roots of 1.

In this paper the quasitriangularlity of Hopf algebras $U_q(\frak g)$
will be studied for $\frak g = sl_2$. It is studied over $\Bbb
C[q,q^{-1}]$ for general values of $q$ and when $q$ is a root of 1.

The main result is that for algebras $U_q(\frak g)$ we have a somewhat
more general property than quasitriangularilty (1)--(3).

\bigskip
{\bf Definition 2.} A Hopf algebra $A$ is called autoquasitriangular
if there exists an automorphism $R$ of $A\otimes A$ (or of an
appropriate completion of $A\otimes A$) distinct from
$\sigma: a\otimes b\mapsto b\otimes a$ such that
$$
\Delta'(a) = R(\Delta(a))    \tag 4
$$
$$
(\Delta\otimes{\op{id}})\circ R =
R_{13}\circ R_{23}\circ (\Delta\otimes{\op{id}})       \tag 5
$$
$$
({\op{id}}\otimes\Delta)\circ R = R_{13}\circ R_{12}\circ
({\op{id}}\otimes\Delta)    \tag 6
$$
Here $R_{12},R_{13},R_{23}$ are automorphisms of
$A\otimes A\otimes A$ such that $R_{12}=R\otimes{\op{id}}$, \
$R_{23}={\op{id}}\otimes R$, \ $R_{13}=(\sigma\otimes{\op{id}})\circ
({\op{id}}\otimes R)\circ(\sigma\otimes{\op{id}})$.

\bigskip
It follows from this definition that the automorphism $R$ satisfies
the  Yang-Baxter equation in End$(A^{\otimes 3})$:
$$
R_{12}\circ R\circ R_{23}\circ R_{23} = R_{23}\circ R_{13}\circ R_{12}  \tag 7
$$
Let $R^{(0)}$ above be an exterior automorphism  of $A\otimes A$
and $R^{(1)}\in A\otimes A$ be an invertible element. Consider the
automorphism
$$
a\mapsto R^{(1)}aR^{(1)^{-1}} \ , \qquad
a,R^{(1)}\in A\otimes A      \tag 8
$$

\bigskip
{\bf Definition 3.} The element $R^{(1)}$ is a {\it universal
$R$-matrix} of the autoquasitriangular Hopf algebra
$(A,R^{(0)},R^{(1)})$ if
\roster
\item"(i)"
$$
\Delta'(a)=R^{(0)}(R^{(1)}\Delta(a)R^{(1)})^{-1}   \tag 9
$$
\item"(ii)" (5) and (6) hold for $R^{(0)}$.
\item"(iii)"
$$
\aligned
(\Delta\otimes{\op{id}})(R^{(1)}) & =
R_{23}^{(0)^{-1}}(R_{13}^{(1)})R^{(1)}_{23}\\
({\op{id}}\otimes\Delta)(R^{(1)} & =
R_{12}^{(0)^{-1}}(R_{13}^{(1)})R_{12}^{(1)}
\endaligned      \tag 10
$$
\endroster
Now for the universal $R$-matrix we will have the following relations:
$$
\aligned
& (R_{12}^{(0)^{-1}}\circ R_{13}^{(0)^{-1}})(R_{23}^{(1)})\cdot
 R_{12}^{(0)^{-1}}(R_{13}^{(1)})\cdot R_{12}^{(1)} = \\
& = R^{(1)}_{12} R_{23}^{(0)^{-1}}(R_{13}^{(1)})R_{23}^{(1)} \ ,
\endaligned     \tag 11
$$
$$
\aligned
& (R_{23}^{(0)^{-1}}\circ R_{13}^{(1)^{-1}})\cdot
   R_{23}^{(0)^{-1}}(R_{13}^{(1)})\cdot R_{23}^{(1)} = \\
& = R_{23}^{(1)}\cdot R_{12}^{(0)^{-1}}(R_{13}^{(1)})\cdot
    R_{12}^{(1)}
\endaligned        \tag 12
$$
We will say that the decomposition
$$
R(a) = R^{(0)}(R^{(1)}aR^{(1)^{-1}})    \tag 13
$$
is a {\it regular splitting} of $R$ if
$$
\aligned
(R_{23}^{(0)^{-1}}\circ R_{13}^{(0)^{-1}})(R_{12}^{(1)}) & =
  R_{12}^{(1)} \ , \\
(R_{12}^{(0)^{-1}}\circ R_{13}^{(0)^{-1}})(R_{23}^{(1)}) & =
  R_{23}^{(1)}
\endaligned            \tag 14
$$
If $R^{(0)}$ and $R^{(1)}$ is a regular splitting of $R$ we have the
following relation for $R^{(1)}$:
$$
R_{12}^{(1)}R_{23}^{(0)^{-1}}(R_{13}^{(1)})R_{23}^{(1)} =
R_{23}^{(1)}R_{23}^{(0)}(R_{13}^{(1)})R_{12}^{(1)}   \tag 15
$$

In the first section we recall the definition and the main properties
of a formal deformation $U_hsl_2$ of $U(sl_2)$. This section also contains the
description of the ``algebraic'' quantum group $U_q(sl_2)$ and its
main properties including autoquasitriangularity. In the second
section the algebra $U_q(sl_2)$ and its relation to $U(sl_2)$
are studied further. The third section contains facts about quantum
$sl_2$ at roots of 1 and the description of autoquasitriangularity of
this quantum group.

The corresponding properties of $U_q(\frak g)$ for simple Lie
algebras $\frak g$ as well as an explicit description of the center
for $U_q(\frak g)$ when $q$ is a root of 1 will be done in a
forthcoming paper.

\bigskip
This work was finished during a visit by the author to the School of
Mathematics and School of Theoretical Physics at the Australian
National University at Canberra.  I would like to thank R. Baxter and
V. Bazhanov for their hospitality. It is also my pleasure
to thank R.Baxter,V. Bazhanov and I. Frenkel for valuable discussions.

\bigskip\bigskip
\centerline{\bf $\S$1. Quantum $sl_2$ over $\Bbb C[[h]]$
  and $\Bbb C[q,q^{-1}]$}

\bigskip
{\bf 1.1} \ Here we recall the definition and the main properties of
the algebra which is called quantum $sl_2$ and is denoted as $U_h
sl_2$.  This is a $\Bbb C[[h]]$-algebra generated by elements
$H, X, Y$ with the following determining relations:
$$
[H,X] = 2X, \qquad [H,Y] =-2Y \ ,   \tag 1.1.1
$$
$$
[X,Y] = \frac{sh\left( \frac{hH}{2}\right)}{sh\left( \frac h2\right)}
\ .
     \tag 1.1.2
$$
This algebra is a Hopf algebra with the comultiplication
$$
\Delta H = H\otimes 1+1\otimes H \ ,    \tag 1.1.3
$$
$$
\Delta X = X\otimes e^{hH/4} + e^{-hH/4}\otimes X \ ,
   \tag 1.1.4
$$
$$
\Delta Y = Y\otimes e^{hH/4}+e^{-hH/4}\otimes Y \ .  \tag 1.1.5
$$
The algebra $U_hsl_2$ is a formal deformation of $Usl_2$:
$$
U_hsl_2/hsU_hsl_2 \simeq Usl_2      \tag 1.1.6
$$
The algebra $U_hsl_2$ is quasitriangular with
$R\in U_hsl_2\hat\otimes U_h sl_2$ (where $\hat\otimes$ is the
$h$-adic completion of $U_h sl_2^{\otimes 2}$) given as follows:
$$
R = {\op{exp}} \left( \frac{H\otimes H}{4}\right) \sum_{n\geq 0}
\frac{(e^h-1)^{2k}}{(e^{hk}-1)(e^{h(k-1)}-1)\dots (e^h-1)} \
e^{\frac h2 \ n(n+1)} (e^{\frac{hH}{4}} X)^k \otimes
(e^{-\frac{hH}{4}} Y)^k      \tag 1.1.7
$$
The representation theory of the algebra $U_hsl_2$ is the same as
the representation theory of $Usl_2[[h]]$ due to the following fact [Dr1][J].

\proclaim{Proposition 1.1.2} There is a homomorphism of algebras
$$
U_hsl_2 {\overset\sim\to\rightarrow} Usl_2[[h]] \ .
$$
Any map $H_h \mapsto H$, \ $X\mapsto X\phi_h(H)$, \
$Y\mapsto Y\phi_h (H)$, where $\phi_h(Z)=\psi_h(Z)=1 \mod h$ \
$(\phi,\psi\in\Bbb C[z][[h]])$ and
$$
\phi_h(z+2)\psi_h(z) = \phi_h(z)\psi_h(z-2) =
\frac{zsh(\frac h2 )}{sh( \frac{hz}{2})}
$$
provides such an isomorphism
\endproclaim

\proclaim{Corollary} ${\op{Center}}(U_hsl_2)\simeq
{\op{Center}}(Usl_2)[[h]]$
\endproclaim

The category $U_hsl_2$-mod of finite-dimensional $U_h \
sl_2$-modules is an abelian category over $\Bbb C[[h]]$. Objects of
this category are pairs $(V,\pi)$, where $V$ is a finite-dimensional
$\Bbb C[[h]]$-vectorspace and ${\overline\kappa}: U_hsl_2\to
{\op{End}}(V)$ is a homomorphism  of algebras.
{\it Morphisms} $f: (V_1,\pi_1)\to (V_2,\pi_2)$ are
$\Bbb C[[h]]$-linear maps from $V_1$ to $V_2$ which are also
$U_h sl_2$-linear: $f\pi_1(a)=\pi_2(a)f$ for each $a\in U_hsl_2$.

The category $U_h sl_2$-mod is a rigid monoidal category with identity
object $1 \!\! 1 \simeq \Bbb C[[h]]$ and with the tensor product
$(V_1,\pi_1)\otimes (V_2,\pi_2)=
(V_1\otimes V_2,(\pi_1\otimes\pi_2)\circ\Delta)$. The object dual to
$(V,\pi)$ is a pair $(V^*,\pi^*\circ S)$ where $V^*$ is a vectorspace
dual to $V$ over $\Bbb C[[h]]$ and $\pi^*(a)$ is a dual
$\Bbb C[[h]]$-linear map to $\pi(a)$. The category $U_hsl_2$-mod is a
strict monoidal category due to coassociativity of the
comultiplication.

The quasitriangularity of the algebra $U_hsl_2$ implies that the
category of $U_hsl_2$-mod is a braided category.  The braiding is a
collection of functorial isomorphisms
$$
c_{VW}: V\otimes W\longrightarrow W\otimes V    \tag 1.1.8
$$
where $c_{VW}=P_{VW}\cdot(\pi_V\otimes\pi_W)(R)$. Here
$P_{VW}(x\otimes y)=y\otimes x$ is a permutation operator, $R$ is the
universal $R$-matrix (1.1.9).

The remarkable property of the deformation $U_hsl_2$ of $Usl_2$ is that
relations between \ exp$\left( \pm\frac{hH}{4}\right)$, \ $X$ and $Y$
are algebraically closed and the action of the comultiplication on
them results in an algebraic combination of these elements. Moreover
relations between these elements and the comultiplication are  defined
over $\Bbb C[e^{\pm h}]$. This observation is a motivation for the
algebra described in the next section.

\bigskip
{\bf 1.2} \ It turns out that the formal deformation described above
also gives a family of Hopf algebras.  The algebra $U_q(sl_2)$
for an undetermined $q$ as the $\Bbb C[q,q^{-1}]$-algebra is generated by
$k,k^{-1}$, $e$ and $f$ with the following determining relations:
$$
kk^{-1} = k^{-1}k = 1 \ , \qquad  ke =qek \ ,    \tag 1.2.1
$$
$$
kf = qf{-1}fk \ , \qquad
ef-fe =\frac{k-k^{-1}}{q-1}   \tag 1.2.2
$$
We will call this algebra polynomial quantum $sl_2$. This is a Hopf
algebra [Dr][J][S1] with the following action of the comultiplication on
generators:
$$
\Delta k = k\otimes k \ ,   \tag 1.2.3
$$
$$
\Delta e = e\otimes k+1\otimes e    \tag 1.2.4
$$
$$
\Delta f = f\otimes 1+k^{-1}\otimes f    \tag 1.2.5
$$

\bigskip
{\bf Remark 1.2.1}  We have the isomorphism of algebras:
$$
U_q(sl_2) {\overset\rightarrow\to\sim} U_{q^{-1}}(sl_2)  \tag 1.2.6
$$
given by a map
$$
q\mapsto q^{-1} \ ,  \qquad
k\mapsto k^{-1} \ ,  \qquad
e\mapsto ek^{-1} \ , \qquad
f\mapsto fk \ .                  \tag 1.2.7
$$

{\bf Remark 1.2.2} Since $U_q(sl_2)$ is defined over $\Bbb
C[q,q^{-1}]$ we can specialize $q$ to any nonzero complex number.
Thus $U_q(sl_2)$ determines a family of Hopf algebras. Monomials
$e^nk^mf^e$ form a linear  basis in $U_q(sl_2)$.  This allows us to
identify these algebras as vectorspaces for different complex values
of $q$. We assume this identification in the rest of this paper.

\bigskip

{\bf 1.2.3} Identify $q=e^h$ and consider $\Bbb C[[h]]$ as a module
over $\Bbb C[q,q^{-1}]$. Then we have an injective homomorphism of
algebras:
$$
\phi: U_q(sl_2) \to U_h sl_2 \otimes_{\Bbb C[[h]]}
  \otimes \Bbb C[[h]][\e ] \ , \qquad \e^2=1    \tag 1.2.8
$$
$$
\phi(k) = e^{\frac{hH}{2}}\otimes\e \ , \quad
\phi(e) = e^{\frac{hH}{4}}X\otimes e \ , \quad
\phi(f) = e^{-\frac{hH}{4}}Y\otimes 1 \ .   \tag 1.2.9
$$

The following seems well known.

\proclaim{Proposition 1.2.4} The center of $U_q(sl_2)$ is generated by
the element
$$
c = ef + \frac{k+qk^{-1}}{(q-1)^2}      \tag 1.2.10
$$
\endproclaim
The central element (1.2.10) was first constructed in [S2]. The fact
that it generates the center of $U_q(sl_2)$ can be found in [DK].

\bigskip
{\bf 1.3} \  The algebra $U_q(sl_2)$ is not quasitriangular in a sense of
definition 1.  But, as we will see, it is autoquasitriangular in the
sense of Definition 2.

Consider the algebra $U_q(sl_2)$ over $\Bbb C[q^{\frac 12},q^{-\frac
12}]$. Define the automorphism $R_0$ of $U_q(sl_2)^{\otimes 2}$
as follows:
$$
R_0(k\otimes 1)=k\otimes 1 \ , \qquad
R_0(1\otimes k)=1\otimes k             \tag 1.3.1
$$
$$
R_0(e\otimes 1)=e\otimes k \ , \qquad
R_0(1\otimes e)=k\otimes e             \tag 1.3.2
$$
$$
R_0(f\otimes 1)=f\otimes k^{-1} \ , \qquad
R_0(1\otimes f)=k^{-1}\otimes f             \tag 1.3.3
$$
Consider the following completions of $U_q(sl_2)$:
$$U_q(sl_2)^{(e)} = \left\{ \sum^\infty_{n=0}
e^n P_n(k^{\pm 1},f)\right\}      \tag 1.3.4
$$
$$
U_q(sl_2)^{(f)} = \left\{ \sum^\infty_{n=0}
Q_n(k^{\pm 1},f) \right\}     \tag 1.3.5
$$
Here $P_n$ and $Q_n$ are polynomials.  It is clear that these
completions of vectorspaces are indeed completions of algebras.

Consider the following completion of $U_q(sl_2)^{\otimes 2}$:
$$
U_q(sl_2)^{\hat\otimes 2} =
\left\{ \sum^\infty_{n=0}
e^n P_n(k^{\pm 1},f)\otimes Q_n(e,k^{\pm 1})f^n\right\}    \tag 1.3.6
$$
Again, this is a completion of the algebra $U_q(sl_2)^{\otimes 2}$,
such that
$$
U_q(sl_2)^{\hat\otimes 2} \hookleftarrow U_q(sl_2)^{(e)}
                               \otimes U_q(sl_2)^{(f)} \ .  \tag 1.3.7
$$
Clearly the element
$$
R_1 = \sum^\infty_{n=0}
\frac{(q-1)^k}{(k)_q!}  \qquad
q^{\frac{k(k-1)}{2}} \ e^k\otimes f^k     \tag 1.3.8
$$
which is the last factor in the $R$-matrix belongs to
$U_q(sl_2)^{\hat\otimes 2}$. Here $(k)_q!=(k)_q\dots(1)_q$,
$(k)_q= \frac{q^k-1}{q-1}$.  It is also clear that (1.3.8) is
invertible in $U_q(sl_2)^{\hat\otimes 2}$.

Consider $U_q(sl_2)^{\otimes 2}$ as a subspace in
$U_q(sl_2)^{\hat\otimes 2}$.

\proclaim{Proposition 1.3.1} The  Hopf algebra $U_q(sl_2)$ is
autoquasitriangular with
$$
R(a) = R_0(R_1 a R_1^{-1})   \tag 1.2.14
$$
where $R_0$ and $R_1$ are defined above in (1.3.1)--(1.3.3) and
(1.3.8). The element $R_1$ is a universal $R$-matrix for $U_q(sl_2)$
(in the sense of Definition 3).
\endproclaim

{\smc Proof.} First consider the algebra $U_h sl_2$ from section 1.1.
Denote
$$
R_0  = {\op{exp}} (\frac h4 H\otimes H) \  ,   \tag 1.3.10
$$
$$
R_1  = \sum_{n\geq 0} \frac{(e^h-1)^{2k}}{(e^{hk}-1)\dots(e^h-1)}
   e^{\frac h2 n(n-1)}(e^{\frac{hH}{4}}X)^n \otimes
   (e^{-\frac{hH}{4}}Y)^n            \tag 1.3.11
$$
It is easy to check that
$$
\aligned
R_0(X\otimes 1) & = (X\otimes e^{\frac{hH}{2}})R_0 \ , \qquad
R_0(1\otimes X) = (e^{\frac{hH}{2}}\otimes X)R_0  \\
R_0(Y\otimes 1) & = (Y\otimes e^{-\frac{hH}{2}})R_0 \ , \qquad
R_0(1\otimes Y) = (e^{-\frac{hH}{2}}\otimes Y)R_0
\endaligned        \tag 1.3.12
$$
Therefore the automorphism $a\mapsto R_0 aR_0^{-1}$ can be extended
>from the automorphism of $U_h sl_2^{\otimes 2}$ to the automorphism of
$U_q(sl_2)^{\otimes 2}$ (it will be an exterior automorphism of
$U_q(sl_2)^{\otimes 2}$). Coparing with (1.3.1)--(1.3.3) we identify
it with $R_0$.

Clearly $R_0$ satisfies conditions (5) and (6) and the element
$R_1$ satisfies (10),(11),(15).  Because there is a homomorphism of
Hopf algebras $U_q(sl_2)\to U_h sl_2$, an automorphism
(1.3.1)--(1.3.3) is a preimage of $R_0 aR_0^{-1}$ in
$U_hsl_2^{\otimes 2}$ and the element (1.3.8) is a preimage of
(1.3.11). We expect Proposition 1.2.3 to be true since the
homomorphism is injective.

The other way to prove Proposition 1.2.3 is by easy direct
computation.

It is easy to see that the splitting on $R_0$ and $R_1$ is a regular
splitting in the sense of (15) and therefore $R_1$ satisfies the
twisted Yang-Baxter equation (16). In section 1.4 this equation will
be written explicitly.

The algebra $U_q(sl_2)$ is a deformation of the universal
enveloping algebra of $sl_2$ in the following sense.

Consider the algebra $Usl_2^{(\e)}\simeq Usl_2\otimes_{\Bbb C}\Bbb C[\e]$
where $\e^2 =1$ and introduce the following Hopf algebra structure on
it:
$$
\aligned
\tilde\Delta H & = H\otimes 1+1\otimes X \ , \qquad
\tilde\Delta(\e) = \e\otimes\e \ , \\
\tilde\Delta X & = X\otimes 1+\e\otimes X \ , \\
\tilde\Delta Y & = Y\otimes 1+\e\otimes Y \ ,
\endaligned          \tag 1.3.13
$$

\proclaim{Proposition 1.3.2} There is an isomorphism of Hopf algebras:
$$
U_q(sl_2)/(q-1)U_q(sl_2) {\overset\sim\to\rightarrow} Usl_2^{(\e)}
     \tag 1.3.14
$$
where
$$
\aligned
\e & = k\mod (q-1) \ , \qquad  H=\frac{k^2-1}{q-1} \mod (q-1) \ , \\
 X & = k^{-1}e\mod (q-1) \ , \qquad   Y=f \mod  (q-1)
\ .    \endaligned          \tag 1.3.15
$$
\endproclaim

The proof is clear.

{\bf 1.4.} Consider the function of complex $z$,
$$
(z;q)_\infty = \prod_{n\geq 1} (1-zq^n)  \ .   \tag 1.4.1
$$
We regard it as an element of $\Bbb C[[q]]$. This product converges to
an analytic function of $z$ in any finite part of $\Bbb C$ if $q$ is a
complex number, and $|q| < 1$.

The following identities are well known:
$$
(z;q)_\infty = \sum_{n\geq 0}
\frac{(-1)^n q^{\frac{n(n+1)}{2}}}{(n)_q! (1-q)^n} \ z^n
    \tag 1.4.2
$$
$$
(z;q)^{-1}_\infty = \sum_{n\geq 0} \frac{q^n}{(n)_q!(1-q)^n} \ z^n
     \tag 1.4.3
$$
This implies the following multiplicative  presentation for the
universal $R$-matrix $R_1$:
$$
\aligned
R_1 & = ((e\otimes f)(q^{\frac 12}-q^{-\frac 12})^2;q)_\infty \\
    & = \prod_{n\geq 1} (1-(q^{-\frac 12}-q^{\frac 12})^2 e\otimes f \
    q^n) \ .
\endaligned          \tag 1.4.4
$$

Let us write the twisted Yang-Baxter equation explicitly for $R_1$ in
terms of generators of $U_q(sl_2)$.

It is well known  that
$$
(u+v;q)_\infty = (u;q)_\infty(v;q)_\infty    \tag 1.4.5
$$
if
$$
uv=qvu  \ .         \tag 1.4.6
$$
The identities (10),(11) for the universal $R$-matrix $R_1$ follows
>from (1.4.5) with
$$
u:= (q^{\frac 12}-q^{-\frac 12})^2 e\otimes k\otimes f \ , \qquad
v:= (q^{\frac 12}-q^{-\frac 12})^2 (1\otimes e\otimes f)
     \tag 1.4.7
$$
for (10) and
$$
u:= (q^{\frac 12}-q^{-\frac 12})^2 (e\otimes k^{-1}\otimes f) \ ,
   \qquad
v:= (q^{\frac 12}-q^{-\frac 12})^2 (e\otimes f\otimes 1)    \tag 1.4.8
$$
for (11).

The proof is an easy computation.

\bigskip
{\bf Remark 1.4.1.}
As was noted at the end of the previous section, the element $R_1$
satisfies the twisted Yang-Baxter relation (16). In terms of function
$(z;q)_\infty$ this means:
$$
(F;q)_\infty(K_{+};q)_\infty(E;q)_\infty
= (E;q)_\infty(K_{-};q)_\infty(F;q)_\infty
     \tag 1.4.9
$$
where
$$
\aligned
F & = (q^{\frac 12}-q^{-\frac 12})^2 e\otimes f\otimes 1 \ ,
\qquad
E = (q^{\frac 12}-q^{-\frac 12})^2 1\otimes e\otimes f \\
K_{+} & = (q^{\frac 12}-q^{-\frac 12})^2 e\otimes k\otimes f \ ,
\qquad
K_{-} = (q^{\frac 12}-q^{-\frac 12})^2
   e\otimes k^{-1}\otimes f \ .
\endaligned     \tag 1.4.10
$$
Notice that these elements satisfy relations similar to $U_q(sl_2)$:
$$
\matrix
K_{+}F  = q^{-1}FK_{+} \ , & \qquad
K_{+}E = q EK_{+} \ , \\
K_{-}F  = q FK_{-} \ , &\qquad
K_{-}E = q^{-1}EK_{-} \\
[E,F]  = (K_{+}-K_{-})(1-q^{-1}) & {} \\
K_{+}K_{-} = K_{-}K{+} \ , &\qquad
\endmatrix          \tag 1.4.11
$$
These relations may be regarded as determining relations for
$C_q[GL^*_2]$.The explanation of this fact remains a bit mysterious.
Notice that funcions $(z;q)_\infty$ also appeared in [FK],where they were
interpreted as ``quantum diligarithms''.

\bigskip\bigskip

\centerline{\bf $\S$2. More on quantum $sl_2$ over
   $\Bbb C[q,q^{-1}]$\footnote{See also [L].}}

\bigskip
Let us clarify the relation between $U_q(sl_2)$ and
$Usl_2$.

\bigskip
{\bf 2.1} \ A representation $V$ of $sl_2$ is said to be an integer if
$H$ acts as a diagonalizable element in $V$ and if
Spec$(H|_V)\subset\Bbb Z$.

Define $\dot Usl_2$ as the following completion of $Usl_2$ in the
category of integer modules. It is generated by $P_\ell$, \ $X$, \
$Y$, \ $\ell\in\Bbb Z$ with determining relations
$$
P_\ell P_m = \delta_{\ell m} P_\ell \ , \qquad
P_\ell X = X P_{\ell +2}
$$
$$
\aligned
P_\ell Y & = YP_{\ell -2} \ , \\
XY-YX & = \sum_{\ell\in\Bbb Z} \ell P_\ell \ , \qquad
     1= \sum_{\ell\in\Bbb Z} P_\ell
\endaligned                               \tag 2.1.1
$$
As a vectorspace $\dot Usl_2$ consists of elements
$\sum_{\ell,m\in\Bbb Z} P_\ell a_\ell(X,Y)$ where $A_\ell(X,Y)$ is a
polynomial over $X,Y$.

The algebra $\dot Usl_2$ is a topological Hopf algebra with the
comultiplication $\Delta: \dot Usl_2\to \dot Usl_2\otimes\dot Usl_2$
acting as:
$$
\aligned
\Delta(X) & = X\otimes 1+1\otimes X \ , \\
\Delta(Y) & = Y\otimes 1+1\otimes Y \ , \\
\Delta P_{\ell} & = \oplus_{\Sb m+n=\ell\\n,m\in\Bbb Z\endSb}
   P_n\otimes P_m
\endaligned            \tag 2.1.2
$$
Define the Hopf algebra $\dot Usl_2^{(\e)}$ as the algebra which is
isomorphic to $\dot Usl_2\otimes_{\Bbb C} \Bbb C[\e^{\frac 12}]/
\e^2=1$.  The comultiplication
$$
\aligned
\Delta P_\ell & =\sum_{\Sb n+m=\ell\\n,m\in\Bbb Z\endSb} P_n\otimes P_m  \ , \\
\Delta X & = X\otimes 1+\e \otimes X \ , \\
\Delta Y & = Y\otimes 1+\e\otimes Y \ ,\\
\Delta\e^{\frac 12} & = Y\otimes 1+\e^{\frac 12}\otimes\e^{\frac 12}
\endaligned          \tag 2.1.3
$$
provides $\dot Usl_2^{(\e)}$ with the Hopf algebra structure
$$
\aligned
\Delta(X) & = X\otimes 1+1\otimes X \ , \\
\Delta(Y) & = Y\otimes 1+1\otimes Y \ , \\
\Delta(P_\ell) & = \oplus_{\Sb m+n=\ell\\n,m\in\Bbb Z\endSb}
P_n\otimes P_m
\endaligned      \tag 2.1.4
$$
Define the Hopf algebra $\dot Usl_2^{(\e)}$ as the algebra which is
isomorphic to $\dot Usl_2\otimes_{\Bbb C}[\e^{\frac 12}]/ \e^2=1$. The
comultiplication
$$
\aligned
\Delta P_\ell & = \sum_{\Sb n+m=\ell\\n,m\in\Bbb Z\endSb}
P_n\otimes P_m \\
\Delta X & = X\otimes 1+\e \otimes X \ ,  \\
\Delta Y & = Y\otimes 1+\e \otimes Y \ , \\
\tilde\Delta \e^{\frac 12} & = \e^{\frac 12}\otimes \e^{\frac 12} \ ,
\endaligned               \tag 2.1.5
$$
provides  $\dot Usl_2^{(\e)}$ with the Hopf algebra structure.

Let $\Delta$ be the usual diagonal comultiplication:
$$
\aligned
\tilde\Delta X & = X\otimes 1+1\otimes X \\
\tilde\Delta Y & = Y\otimes 1+1\otimes Y  \\
\tilde\Delta P_\ell & = \sum_{\Sb n+m=\ell\\n,m\in\Bbb Z\endSb}
   P_n\otimes P_m \\
\Delta\e^{\frac 12} & = \e^{\frac 12}\otimes \e^{\frac 12} \
\endaligned               \tag 2.1.6
$$
which also provides a Hopf algebra structure on
$\dot Usl_2\otimes_{\Bbb C} \Bbb C[\e^{\frac 12}]$.

Consider the following element in $\dot Usl_2^{(\e)\otimes 2}$:
$$
F=\oplus_{n\in\Bbb Z} \e^{\frac n2} \otimes P_n    \tag 2.1.7
$$

\proclaim{Proposition 2.1.1} (1) The comultiplication (2.1.3) is related
to the diagonal comultiplication by twisting with the element $F$:
$$
\Delta (a) = F\cdot\tilde\Delta (a)\cdot F^{-1}   \tag 2.1.8
$$

(2) The element $F$ has the following properties:
$$
\aligned
(\Delta\otimes{\op{id}})(F) = F_{13}F_{23} \\
({\op{id}}\otimes\Delta)(F) = F_{13}F_{12}
\endaligned                \tag 2.1.9
$$
\endproclaim

The proof is by straightforward computation.

\proclaim{Corollary 2.1.2} The algebra $\dot U sl_2^{(\e)}$ with the
comultiplication (2.1.3) is a quasitriangular Hopf algebra with
$$
R = \sigma (F)F^{-1} = \sum_{nm\in\Bbb Z}
\e^{-\frac m2} P_n\otimes \e^{\frac n2} P_m     \tag 2.1.10
$$
where $\sigma(a\otimes b)=b\otimes a$.
\endproclaim

\bigskip
{\bf 2.2} Define the algebra $\dot U sl_2^{(\e)}$ as the
$\Bbb C[q^{\frac 12},q^{-\frac 12}]$-algebra generated by $e$, \
$f$, \ $\e^{\frac 12}$ and $P_\ell$ with the following determining
relations:
$$
\aligned
P_\ell P_m = \delta_{\ell m} P_\ell \ , \qquad
q = \sum_{\ell\in\Bbb Z} P_\ell \\
P_\ell e = eP_{\ell+2} \ , \qquad
P_\ell f = fP_{\ell-2} \\
ef-fe = \e\sum_{\ell\in\Bbb Z} (\ell)_{q^2} q^{-\ell} (q+1)P_{\ell}  \\
\e^2 = 1 \ .
\endaligned          \tag 2.2.1
$$
As a vectorspace $\dot U sl_2^{(\e)}$  consists of sums
$\sum_{\ell\in\Bbb Z} P_\ell a_\ell(e,f,\e^{\frac 12})$ where $a_\ell$
are polynomials.

The following comultiplication provides $\dot U sl_2^{(\e)}$ with the
Hopf algebra structure:
$$
\aligned
\Delta P_\ell & = \oplus_{n+m=\ell} P_n\otimes P_m \\
\Delta e & = \sum_{\ell\in\Bbb Z}e\otimes q^{\ell /2}\e
   P_\ell+1\otimes e  \\
\Delta f & = f\otimes 1+\sum_{\ell\in\Bbb Z} q^{-\frac 12}
   \e P_\ell\otimes f \\
\Delta\e^{\frac 12} & = \e^{\frac 12}\otimes\e^{\frac 12}
\endaligned              \tag 2.2.2
$$

\proclaim{Theorem 2.2.1} The algebra $\dot U sl_2^{(\e)}$ is
quasitriangular with
$$
R  = \left( \sum_{m,n\in\Bbb Z} q^{\frac{nm}{4}}
      \e^{-\frac n2} P_m\otimes \e^{\frac m2} P_n\right) \cdot
   \cdot \sum_{k\geq 0} \frac{(q-1)^k}{(k)_q!} \
    q^{\frac{k(k-1)}{2}} e^k\otimes f^k                \tag 2.2.3
$$
\endproclaim

Notice that element $\e^{\frac 12} -1$ generates the Hopf ideal.
Let $\dot U_q(sl_2)$ be the corresponding quotient algebra
$$
\dot U_q(sl_2) = \dot U_q^{(\e)}(sl_2)/ \< \e^{\frac 12}-1\>   \tag 2.2.4
$$

{\bf Remark 2.2.2.} It is clear that quasitriangular Hopf algebra
$(\dot U_q^{(\e)}(sl_2),\Delta, R)$ is a deformation of Hopf algebra
$(U^{(\e)}(sl_2),\Delta, R)$ described in the previous section:
$$
\dot U_q^{(\e)}(sl_2)[[q-1]]/(q-1)U_q^{(\e)}(sl_2)[[q-1]]
\simeq U^{(\e)}(sl_2)
$$

\proclaim{Proposition 2.2.3} (1) There is an isomorphism of algebras\newline
$\phi$: \ $\dot U_q(sl_2)^{(\e)}{\overset\sim\to\rightarrow}
\dot U_q(sl_2)\otimes_{\Bbb C} \Bbb C[\e^{\frac 12}]$,
$$
\phi(P_\ell)  = P_\ell\otimes 1 \ , \qquad
\phi(e) = e\otimes\e \ ,  \qquad
\phi(f)  = f\otimes 1                   \tag 2.2.5
$$

(2) The map $\tilde\Delta : \dot U_q(sl_2)^{(\e)}\to
U_q(sl_2)^{(\e)^{\otimes 2}}$,
$$
\aligned
\tilde\Delta P_\ell & = \oplus_{n+m=\ell} P_n\otimes P_m \ , \\
\tilde\Delta e & = \sum_{\ell\in\Bbb Z} e\otimes q^{\ell/2} P_\ell
   +1\otimes e  \ , \\
\tilde\Delta f & = f\otimes 1 +\sum_{\ell\in\Bbb Z}
   q\otimes q^{\ell/2} P_\ell \otimes f \ , \\
\tilde\Delta\e^{\frac 12} & = \e^{\frac 12} \otimes \e^{\frac 12}
\endaligned          \tag 2.2.6
$$
determines a Hopf algebra structure on $\dot Uq(sl_2)^{(\e)}$ and it
is related to the comultiplication (2.2.2) by the twist
$$
\tilde\Delta (\phi (a)) = F^{-1}( \phi\otimes\phi)
(\Delta (a)) F   \tag 2.2.7
$$
where $F$ is the element (2.1.5)

(3) The Hopf algebra $(\dot Uq(sl_2)[\e^{\frac 12}] \tilde\Delta)$
is quasitriangular with
$$
\tilde R  =\left( \sum_{n,m\in\Bbb Z} q^{\frac{nm}{4}}\cdot
   P_m\otimes P_n\right) \sum_{k\geq 0}
   \frac{(q-1)^k}{(k)_q!}
  \cdot  q^{\frac{k(k-1)}{2}} e^k\otimes f^k        \tag 2.2.8
$$
\endproclaim

{\smc Proof.} Statements (1) and (2) are an easy straightforward
exercise. The statement (3d) follows from general facts about
twistings of quasitraingular Hopf algebras [Dr]. For the twisted
$R$-matrix we have:
$$
R = F_{21}\tilde R F^{-1}  \tag 2.2.9
$$
Formula (2.2.8) follows from relations:
$$
\aligned
F_{21} P_n\otimes P_m F^{-1} & =
   \e^{-\frac m2} P_n\otimes P_m \e^{\frac n2} \\
F_{21} e^k\otimes f^k F^{-1} & = (\e e)^k \otimes f^k
\endaligned         \tag 2.2.10
$$

\bigskip
{\bf Remark 2.2.4.} The quasitriangular Hopf algebra $\dot U_q sl_2$
is a deformation of $\dot Usl_2$:\newline
$\dot Uq(sl_2)[[q-1]]/(q-1)\dot Uq(sl_2)[[q-1]] \simeq \dot Usl_2$.

For more on algebras $\dot Uq(\frak G)$, see [L].

\bigskip\bigskip\bigskip

\subheading{$\S$3. Quantum $sl_2$ at roots of 1}

{\bf 3.1.} Let $\e$ be a primitive root of 1 of degree $\ell$ and let
$\ell$ be odd.

Let $q=e^h\e$ and consider
$$
U_\e(sl_2) = U_q(sl_2)[[h]]/hU_q(sl_2)[[h]]  \tag 3.1.1
$$
If $A$ is an algebra we denote by $Z(A)$ its center.

\proclaim{Proposition 3.1.1} (1) Elements $e^\ell,k^\ell,f^\ell$
belong to the center of $U_\e(sl_2)$.

(2) The center of $U_\e(sl_2)$ is generated by $e^\ell,k^\ell,f^\ell$
and by
$$
c = ef + \frac{k+k^{-1}\e}{(\e -1)^2}    \tag 3.1.2
$$
freely modulo relation
$$
{\op{det}}\left(
{\undersetbrace{\ell}\to
{\matrix
  (\e -1)^2c  &  \e      & \circ\dots & \circ \ 1 \\
            1 &  \ddots  &  O    & \circ \\
      \circ   &   {}     &  {}   & {} \\
      \vdots  &   \ddots &       & \vdots \\
      \circ   &    O     &  {}   &    \e  \\
      \e      &   \circ\dots\circ  &  1    & c(\e -1)^2
\endmatrix}}   \right) =
{(\e -1)^{2\ell} e^\ell f^\ell+k^\ell+k^{-\ell}}-2     \tag 3.1.3
$$

(3) The algebra $U_\e(sl_2)$ is finite-dimensional over
$Z(U_\e(sl_2))$.
\endproclaim

See [DK] for details.

One can introduce the Poisson structure on the center of
$U_\e(sl_2)$ according to the following general construction.

Let $A_h$ be an algebra deformation of associative algebra $A$.
We assume that $A_h=A[[h]]$. We denote the multiplication in $A_h$ by
$m_h: A[[h]]^{\otimes 2}\to A[[h]]$ and we have $m_h=m$ mod $h$ where
$m$ is the multiplication in $A$.

\proclaim{Proposition 3.1.2} Let $a,b\in A[[h]]$ and either
$a {\op{ \ mod \  }}h$ or $b {\op{ \ mod  \ }}h$ belongs to the center of $A$.
Then the element
$$
\{a {\op{ \ mod  \ }}h,b {\op{ \ mod \ }}h\} =
\frac 1h(m_h(a,b)-m_a(b,a)){\op{ \ mod \ }} h    \tag 3.1.4
$$
is defined and $\{\cdot , \cdot\}$ determines a Poisson
structure on the algebra $Z(A)$, and $\{\cdot ,\cdot\}$
determines a Poisson action of the Poisson algebra
$(Z(A),\{\cdot ,\cdot \})$ by derivations of $A$.
\endproclaim

Proof is straightforward (see for example [DP]). Notice
that $\{\cdot ,\cdot \}$ may be identically zero.

In our case $A$ is $U_\e(sl_2)$ and $A_h$ is $U_{\e e^h}(sl_2)$ and the
Poisson structure $\{\cdot, \cdot \}$ can be computed
explicitly between generators of $Z(U_\e(sl_2))$. The answer is:
$$
\aligned
\{ k^\ell,e^\ell\} & = \ell^2 k^\ell e^\ell, \qquad
\{ k^\ell,f^\ell\} = -\ell^2 k^\ell f^\ell \ , \\
\{ e^\ell,f^\ell\} & = \left( \frac{1}{\e -1} \right)^{2\ell +1}\cdot
   \ell^2\cdot (k^\ell-k^{-\ell}) \\
\{ c,a\} & = 0
\endaligned             \tag 3.1.5
$$
for each $a\in Z(U_\e(sl_2))$.

The Poisson action of $Z(U_\e(sl_2))$ on $U_\e(sl_2)$ can also be
computed explicitly:
$$
\aligned
\{ k^\ell,e\} & = \ell ek^\ell \ , \qquad
\{ k^\ell,f\}   = -\ell f k^\ell \ , \\
\{ k,e^\ell\} & = \ell k  \ell^\ell \ , \qquad
\{ e^\ell,f\}   = \frac{\ell}{\e-1} \ e^{\ell-1}\cdot
   \frac{k-k^{-1}\e}{\e -1} \ , \\
\{ k,f^\ell\} & = -\ell kf^\ell \ , \qquad
\{ e,f^\ell\}  = \frac{\ell}{\e -1}\cdot
      \frac{k-k^{-1}\e}{\e-1} \cdot f^{\ell -1} \ , \\
\{ c,e\} & =\{ c,k\} = \{ c,f\} = 0
\endaligned                \tag 3.1.6
$$
All these formulas can be easily derived from relations in $U_q(sl_2)$
and from the identity
$$
\aligned
e^\ell f^\ell & = \sum^\ell_{m=0}
   \frac{(n)_q!}{(n-\ell+m)_q!} \
   \frac{kq^{m-\ell +1}-k^{-1}q^{-m+\ell -n}}{q-1} \ \dots \\
  & \dots \frac{k-k^{-1}q^{\ell-m-n}}{q-1} \cdot f^{n-\ell +m}\cdot
     e^m
\endaligned               \tag 3.1.7
$$
which can be found in [K].

Denote by $Z_0(U_\e(sl_2))$ the central subalgebra in $U_\e(sl_2)$
generated by $e^\ell,f^\ell,k^{\pm\ell}$.

\bigskip
{\bf Remark 3.1.3.} The algebra $Z(U_\e(sl_2))$ is finite-dimensional
over $Z_0(U_\e(sl_2))$.

\proclaim{Proposition 3.1.4} The subalgebra $Z_0(U_\e(sl_2))$ is a Hopf
subalgebra with the comultiplication
$$
\aligned
\Delta k^\ell & = k^\ell\otimes k^\ell \ ,  \\
\Delta e^\ell & = e^\ell\otimes k^\ell + 1\otimes e^\ell  \ , \\
\Delta f^\ell & = f^\ell\otimes 1 + k^{-\ell}\otimes f^\ell  \ .
\endaligned      \tag 3.1.8
$$
\endproclaim
The proof is an elementary corollary of the identities
$$
\aligned
\Delta f^n & = \sum^n_{s=0} \left( \matrix n\\s \endmatrix \right)_q
   f^s k^{-n+s} \otimes f^{n-s}  \ , \\
\Delta e^n & = \sum^n_{s=0} \left( \matrix n\\s \endmatrix \right)_q
   e^s\otimes e^{n-s} k^s \ , \\
\Delta k^n & = k^n\otimes k^n
\endaligned                \tag 3.1.9
$$

The following is a general fact about Hopf algebra deformations.
Let $A$ be a Hopf algebra with multiplication $m$ and comultiplication
$\Delta$.  Let $A_h$ be a Hopf algebra deformation of $A$ such that
$A_h = A[[h]]$ as a vectorspace, $\Delta_h=\Delta$ and
$m_h=m$ mod $h$.

\proclaim{Proposition 3.1.5} Let $X(A)\subset Z(A)$ be a central
subalgebra which is a Hopf subalgebra and let
$\{\cdot ,\cdot \}$ be a Poisson structure (3.1.4) on
$X(A)$. Then $X(A)$ is a Hopf-Poisson algebra in the sense of [Dr]:
$$
\Delta (\{ a,b,\}) = \{\Delta (a),\Delta (b)\}   \tag 3.1.10
$$
where $\{ a\otimes b,c\otimes d\}=\{ a,c\}\otimes bd +
ac\otimes \{ b,d\}$.
\endproclaim

{\smc Proof.} Let us compute $\Delta (\{ a,b\})$:
$$
\aligned
\Delta (\{ a,b\}) & = \frac 1h \Delta(m_h(a,b)-m_h(b,a))\mod h\\
  & = \frac 1h (m_h(a^{(1)},b^{(1)})m_h(a^{(2)},b^{(2)})
               -m_h(b^{(1)},a^{(1)})m_h(b^{(2)},a^{(2)}))\mod h \\
  & = \{ a^{(1)},b^{(1)}\}m(a^{(2)},b^{(2)})
      + m(a^{(1)},b^{(1)})\{ a^{(2)},b^{(2)}\} \\
  & = \{ \Delta (a),\Delta (b)\}
\endaligned
$$
This proves the proposition.

\bigskip
It is  known that the coalgebra structure on $U_q(sl_2)$, \
$q\in\Bbb C^*$ does not depend on $q$ \ [FRT] and therefore we have

\proclaim{Corollary 3.1.6} The central subalgebra $Z_0(U_\e(sl_2))$ is
a Hopf-Poisson algebra with the comultiplication (3.1.8) and with
Poisson structure (3.1.5).
\endproclaim

In the case of quantum $sl_2$ it is also an easy explicit computation
which proves that (3.1.8) is compatible with Poisson brackets (3.1.5).

\bigskip
{\bf 3.2.} Let $\frak A$ be a Lie algebra with Lie bracket
$[ \cdot , \cdot ]$. Denote by $H(x,y)_{[\cdot,\cdot]}$
its Campbell-Hausdorff series; so that
$$
\exp(x)\cdot\exp(y) = \exp(H(x,y)_{[ \cdot \ , \ \cdot \ ]})
     \tag 3.2.1
$$
$$
H(x,y)_{[ \cdot \ , \ \cdot \ ]} = x+y+\tfrac 12 [x,y]+\dots
$$
Here we assumed $\frak A\hookrightarrow U\frak A$ and the
multiplication in the left side is taken in $U\frak A$. If $\frak A$
is infinite-dimensional $U\frak A$ should be properly defined as a
topological algebra.

Another important fact is the identity
$$
\exp(x)\cdot y\cdot\exp(-x) = \sum^\infty_{n=0}
\frac{1}{n!} {\undersetbrace{n}\to{[x[\dots[x }}   ,y]\dots]  \tag 3.2.2
$$
which holds in $U\frak A$ (or in appropriate topological algebra).

Suppose $(A,\{ \cdot ,\cdot\})$ is a Poisson algebra and $A_h$
is an associative algebra which is a formal deformation of
$(A,\{ \cdot , \cdot\})$. Assume the identification of vectorspaces
$A_h=A[[h]]$. Denote the multiplication in $A_h$ as
$m_h: A_h^{\otimes 2}\to A_h$. We have
$$
m_h(a,b) = ab + \frac h2 \{ a,b\} + o(h^2)    \tag 3.2.3
$$
where $(a,b)\mapsto ab$ is the commutative multiplication in $A[[h]]$.

Introduce the following Lie algebra structure on $A_h[h^{-1}]$:
$$
[x,y] = \frac 1h (m_h(x,y)-m_h(y,x))  \ .      \tag 3.2.4
$$
For $x\in A[[h]]$ define $x_0=x$ mod $h$, \ $x_0\in A$. Clearly
$$
[x,y]_0 = \{ x_0,y_0\}      \tag 3.2.5
$$
where $x,y\in A[[h]] \hookrightarrow A[h^{-1},h]$.

In order to define Campbell-Hausdorff series for the Lie algebra
$(A_h,[ \ \cdot \ , \ \cdot \ ])$ we have to consider an appropriate
completion of $A$ when $A$ is infinite-dimensional. For example we can
assume that $A_h$ is a filtered
algebra with filtration $\{ A_h^{(n)}\}$ and then
$$
[A^{(n)},A^{(m)}] \subset A^{(n+m-1)}    \ . \tag 3.2.6
$$
In this case $\hat A_h$ will be the completion of $A_h$ with respect to this
filtration.

\proclaim{Lemma 3.2.1} Let $x,y\in A[[h]] \hookrightarrow
A[[h^{-1},h]]\hookrightarrow \hat A[[h^{-1},h]]$. Then the product
$$
m_h(\exp (\frac xh ),\exp(\frac yh )) =
    \exp\left( \frac{H(x,y)_{[ \ \cdot \ , \ \cdot \ ]}}{h}\right)
                 \tag 3.2.7
$$
is defined in $\hat A[[h^{-1},h]]$. Here $H(x,y)$ is the
Campbell-Hausdorff series for Lie algebra $(\hat A_h,[ \  \cdot \ ,
\ \cdot \ ])$.
\endproclaim

Note that (3.2.5) implies
$$
H(x,y)_{[ \  \cdot \ , \ \cdot \ ]} {\op{ \ mod \ }} h =
H(x_0,y_0)_{\{ \ \cdot \ , \ \cdot \ \}}      \tag 3.2.8
$$
where $x,y\in\hat A[[h]]$.

\proclaim{Lemma 3.2.2} For $x,y\in A_h\hookrightarrow A[[h^{-1},h]]$,
the product
$$
m_h(m_h({\op{exp}}(\frac xh ),y),e^{-\frac xh}) =:
{\op{exp}}(x)*y          \tag 3.2.9
$$
is defined over $\hat A[[h^{-1},h]]$ and
${\op{exp}}(x)*y\in\hat A[[h]]\hookrightarrow \hat A[[h^{-1},h]]$.
\endproclaim

This lemma follows immediately from (3.2.2) and from the fact that
$xy\!-\!yx\! =\!0$ mod $h$.

Notice that
$$
{\op{exp}}(x)*y {\op{ \ mod \ }} h = {\op{exp}}(x_0)\circ y_0
                   \tag 3.2.10
$$
where
$$
{\op{exp}}(x_0)\circ y_0  = \sum_{n\geq 0} \frac{1}{n!}
{\undersetbrace{n}\to\{ x_0\{\dots \{ x_0 }}}}  ,y_0\}\dots\}}
            \tag 3.2.11
$$

\bigskip\bigskip
{\bf 3.3} \  Let $SL^*_2$ be a Lie group dual to $SL_2$ in the sense of
vdual Lie-Poisson groups [Dr].  This group may be regarded as a group
of pairs of triangular matrices:
$$
\left( \left( \matrix  1 & e \\  0 & k  \endmatrix \right) \ ,
       \left( \matrix  1 & 0 \\  f  & k^{-1} \endmatrix \right)\right)
  \tag 3.3.1
$$
with pairwise multiplication $(x,y)(x',y')=(xx',yy')$.

This group is a Lie-Poisson group [Dr] which means, in particular,
that the algebra of algebraic function $C[SL^*_2]$ is a Hopf-Poisson
algebra with the comultiplication induced by a group multiplication in
$SL^*_2$ and with the following Poisson brackets
between coordinate functions:
$$
\{ e,f\} = k-k^{-1} \ , \qquad
\{ k,e\} = ke \ , \qquad
\{ k,f\} = -kf                  \tag 3.3.2
$$

{\bf Remark 3.3.1.} We have an isomorphism of coalgebras (see for
example [FRT]):
$$
U_q(sl_2)\otimes_{\Bbb C[q,q^{-1}]} \Bbb C(q) \simeq
C[SL_2^*](q)            \tag 3.3.3
$$

\proclaim{Proposition 3.3.2}  There is an isomorphism of Hopf-Poisson algebras:
$$
\phi : Z_0(U_\e(sl_2)) {\overset\sim\to\longrightarrow}C[SL^*_2]
             \tag 3.3.4
$$
$$
\phi(e^\ell) = (\e -1)^{-\ell-1}e \ , \qquad
\phi(k^\ell)=k \ , \qquad
\phi(f^\ell) = (\e -1)^{-\ell} f         \tag 3.3.5
$$
$$
\phi( \{ a,b\}) = \ell^2\{\phi (a),\phi (b)\}    \tag 3.3.6
$$
\endproclaim

The proof is clear.

Let $C[[SL^*_2]]$ be the algebra of jets of functions on $SL^*_2$ in
the neighborhood of 1. As a vectorspace it consists of the formal
power series over $k-1, \ e, \ f$.

\proclaim{Theorem 3.3.3} The algebra $C[[SL^*_2]]$ is
autoquasitriangular with the automorphism $R$ of the form:
$$
R(a) = \exp(r_0)\circ\exp(r_1)(a)  \ .  \tag 3.3.7
$$
\endproclaim
Here
$$
r_0 = \tfrac 14 z\otimes z       \tag 3.3.8
$$
$$
r_1 = Li_2(e\otimes f) \ , \qquad
       Li_2(x) = -\int^x_0 \frac{\log(1-y)}{y} \ dy    \tag 3.3.9
$$
and we assumed $k =\exp(\frac z2 )$.

{\smc Proof.} Consider the algebra $C_h[[SL^*_2]]$ over
$\Bbb C[[h]]$ generated by $\overline e, \ \overline f, \ z$
(we will assume that it is completed by formal power series with
respect to $\overline e, \ \overline f$ and $z$) with determining
relations:
$$
\aligned
& \overline e\overline f -\overline f\overline e =
( e^{\frac h2}-e^{-\frac h2})(e^{\frac z2}-e^{-\frac z2}) \\
& [z,\overline e] = 2he \ , \qquad
  [z,\overline f] = -2h\overline f
\endaligned          \tag 3.3.10
$$
It is a Hopf algebra with the comultiplication
$$
\aligned
\Delta z & = z\otimes 1+1\otimes z \ , \\
\Delta\overline e & = \overline e \otimes e^{\frac z2} +1
     \otimes \overline e \ , \\
\Delta\overline f & = \overline f \otimes 1+e^{-\frac z2}
    \otimes \overline f \ .
\endaligned          \tag 3.3.11
$$
Clearly this algebra is a Hopf algebra deformation of Hopf-Poisson
algebra $C[[S^*_2]]$.

The map $\phi :U_q(sl_2)\to C_h[[SL^*_2]][h^{-1}]$
$$
\phi (e) = \frac{\overline e}{(e^{\frac h2}-e^{-\frac h2})} \ , \qquad
\phi (f) = \frac{\overline f}{(e^{\frac h2}-e^{-\frac h2})} \ , \qquad
\phi (k) = e^{\frac z2}      \tag 3.3.12
$$
is a homomorphism of algebras. Here we assumed that $q=e^h$ and
considered $U_q(sl_2)$ over $\Bbb C[[h]]$.

Consider the image of $R_1$ under the extension of $\phi\otimes\phi$
to an appropriate completion of $U_q(sl_2)^{\otimes 2}$:
$$
(\phi\otimes\phi)(R_1) =
(\overline e\otimes\overline f; e^h)_\infty     \tag 3.3.13
$$
Here $(z;e^h)_\infty$ is the asymptotics of the function
$(z;q)_\infty$ at $q\to 1$ (see Lemma from section 3.4).
$$
(\overline e\otimes\overline f; e^h) = \exp
\left( -\frac{Li_2(\overline e\otimes\overline f)}{h}\right)
(1-\overline e\otimes\overline f)^{-\frac 12}\cdot
(1+O(h))      \tag 3.3.14
$$
This asymptotics in the element of $\Bbb C_h[[SL^*_2]]^{\hat\otimes
2}[[h^{-1}]]$ where the tensor product completed by power series over
$\overline e\otimes 1, \ 1\otimes\overline e, \dots$.

According to the previous section the element
$$
R^{(1)}(a)=(\phi\otimes\phi)(R_1)\cdot a\cdot
(\phi\otimes\phi)(R_1)^{-1}          \tag 3.3.15
$$
where $a\in C_h[[SL^*_2]]^{\otimes 2}$ exists in
$C_h[[SL^*_2]]^{\hat\otimes 2}$ and determines the algebra
automorphism of $C_h[[SL^*_2]]^{\hat\otimes 2}$.

We will denote by $R^{(0)}$ the image of
the automorphism (1.2.6)--(1.2.8) under $\phi\otimes\phi$:
$$
\aligned
R^{(0)}(\overline e\otimes 1) & = \overline e\otimes e^{\frac z2} \ ,
   \qquad
R^{(0)}(\overline f\otimes 1) = \overline f\otimes e^{-\frac z2} \\
R^{(0)}(1\otimes\overline e) & = e^{\frac z2}\otimes\overline e \ ,
   \qquad
R^{(0)}(1\otimes\overline f) = e^{-\frac z2}\otimes\overline f \\
R^{(0)}(z\otimes 1) & = z\otimes 1 \s , \qquad
R^{(0)}(1\otimes z) = 1\otimes z \ .
\endaligned          \tag 3.3.16
$$

Define
$$
\overline R(a) = R^{(0)}\circ R^{(1)}(a)   \ . \tag  3.3.17
$$
 From the definition of $R$ and from the autoquasitriangularity of
$U_q(sl_2)$ we deduce that $C_h[[SL^*_2]]$is autoquasitriangular with
$\overline R$ defined in (3.3.17).

Moreover, we can represent $R^{(0)}$ as follows:
$$
R^{(0)}(a) = \exp\left( \frac{z\otimes z}{4h}\right)\cdot a\cdot\exp
                \left( -\frac{z\otimes z}{4h}\right)   \tag 3.3.18
$$
As was explained in the previous section this product exists in
$C_h[[SL_2^*]]^{\hat\otimes 2} \hookrightarrow
C_h[[SL_2^*]]^{\hat\otimes 2} [[h^{-1}]]$.

Since the Hopf algebra $C_h[[SL^*_2]]$is a deformation of Hopf-Poisson algebra
$C_h[[SL^*_2]]$ the latter is autoquasitriangular with
$$
R(a{\op{ \ mod  \ }}h) = \overline R(a) {\op{ \ mod  \ }}h   \tag 3.3.19
$$
and since (3.3.14), (3.3.15) and (3.3.18)
$$
R(a) = \exp(r_0)\circ\exp(r_1)(a)    \tag 3.3.20
$$
where exp$(a)\circ b$ is defined in (3.2.11).

Note that in the proof of Theorem 3.3.3 we have also proven an
autoquasitriangularity of $C_h[[SL^*_2]]$.

{\bf Remark 3.3.4} The analog of Theorem 3.3.3 for an arbitrary simple
Lie algebra $\frak G$ with the standard Lie bialgebra structure [Dr]
has been given in [R] where linear terms of $r_0$ and $r_1$ have been
described.

\bigskip\bigskip
{\bf 3.4} Let $(z;q)_\infty$ be the function defined in (1.4.1) for
$|q| < 1$.

\proclaim{Lemma 3.4.1} The function $(z;q)_\infty$ has the following
asymptotics when $q\to\e$, \ $\e^\ell =1$, \ $\ell$-odd.
$$
\aligned
(z;q)_\infty & \to \exp\left( -\frac{1}{\ell^2 h} \int^{z^\ell}_0
\frac{\log(1-t)}{t} \ dt\right) \cdot \\
  & \cdot (1-z^\ell)^{-\frac 12} \prod^\ell_{m=0}
(1-\e^m z)^{-\frac m\ell} (1+O(h))
\endaligned          \tag 3.4.1
$$
where $q=\e e^h$, \ $h\to 0$.
\endproclaim

{\smc Proof.} The function $(z;q)_\infty$ satisfies the following
difference equation
$$
(zq;q)_\infty = \frac{1}{1-zq} \ (z;q)_\infty    \tag 3.4.2
$$
and it is uniquely determined by this property and by the condition
$(0;q)_\infty =1$. Iterating this equation we have:
$$
(zq^\ell ;q)_\infty =\frac{1}{(1-zq)\dots (1-zq^\ell)} \
(z;q)_\infty            \tag 3.4.3
$$
The function (3.4.1) presents the asymptotics of the solution to the
equation (3.4.3) normalized by $(0;q)_\infty =1$.

\bigskip\bigskip
{\bf 3.5.} Let $\e$ be a root of 1 of odd degree $\ell$.

\bigskip
{\bf Definition 3.5.1.} The algebra $\overline{U_\e(sl_2)}$ is a
complex algebra generated by elements
$\overline e, \ \overline f, \ \omega, \ z$ (complicated by formal
power series over $\overline e, \ \overline f, \ z$) with the
following determining relations:
$$
\aligned
\omega\overline e & = \e\overline e\omega \ , \qquad
\omega\overline f = \e^{-1}\overline f\omega \ , \\
\omega^\ell & = 1 \ , \qquad
\overline e\overline f-\overline f\overline e =
\frac{\omega e^{\frac z2}-\omega^{-1}e^{-\frac z2}}{\e -1}
\endaligned             \tag 3.5.1
$$
and $z\in$ center of $\overline{U_\e(sl_2)}$.

Clearly $\overline{U_\e(sl_2)}$ is a Hopf algebra with the
comultiplication
$$
\aligned
\Delta\overline e & = \overline e \otimes \omega e^{\frac z2} +
   1\otimes\overline e \ , \\
\Delta\overline f & = \overline f\otimes 1 +\omega^{-1}
   e^{-\frac z2} \otimes\overline f \ , \\
\Delta z & = z\otimes 1+1\otimes z \ ,
\endaligned             \tag 3.5.2
$$
and the map $\phi :U_\e(sl_2)\to\overline{U_\e(sl_2)}$,
$$
\phi(k) =\omega e^{\frac z2} \ , \qquad
\phi(e) = \overline e \ , \qquad
\phi(f) = \overline f                    \tag 3.5.3
$$
determines a homomorphism of Hopf algebras.

The algebra $\overline{U_\e(sl_2)}$ has all properties absolutely
similar to $U_\e(sl_2)$:
\roster
\item"$\bullet$" the center of $\overline{U_\e(sl_2)}$ is generated by
   $\overline e^\ell, \ \overline f^\ell, \ z$ and by
$$
    c = ef +
    \frac{\omega e^{\frac z2}+\e\omega^{-1}e^{-\frac z2}}{(\e-1)^2}
         \tag 3.5.4
$$
\item"$\bullet$" $Z(\overline{U_\e(sl_2)})$ is generated by
   $\overline e^\ell, \ \overline f^\ell$, \ $z$, \ $c$
   freely modulo the relation (3.1.3) where we have to replace
   $k\to e^{\frac z2}\omega$
\item"$\bullet$" $\overline{U_\e(sl_2)}$ is finite-dimensional
   over $Z(\overline{U_\e(sl_2)})$.
\item"$\bullet$" the central subalgebra $Z_0(\overline{U_\e(sl_2)})$
  is a Hopf subalgebra in $\overline{U_\e(sl_2)}$ and Poisson
  brackets (3.1.5),(3.1.6) determine the structure of
  Hopf-Poisson algebra on $Z_0(\overline{U_\e(sl_2)})$
  together with its Poisson action on $\overline{U_\e(sl_2)}$.
\item"$\bullet$"  there is an isomorphism of Hopf-Poisson algebras
  $Z_0(\overline{U_\e(sl_2)})\simeq C[[SL^*_2]]$.
\endroster

Let $\e^{\frac 14}$ be the $4\ell^{\op{th}}$-root of 1, such that
$\e^{\frac{\ell}{4}} := (\e^{\frac 14})^\ell =i$ and
consider $\overline{U_\e(sl_2)}$
over $\Bbb C[\omega^{\frac 14},\omega^{-\frac 14}]$. Then we
have:

\proclaim{Theorem 3.5.2} (1) The algebra $\overline{U_\e(sl_2)}$ is
autoquasitriangular with
$$
R(a) = R^{(0)}(R^{(1)}a R^{(1)^{-1}})       \tag 3.5.5
$$
where
$$
R^{(0)}(a) = \exp(r_0)\circ\exp(r_1)    \tag 3.5.6
$$
$$
r_0 = \tfrac 14 z\otimes z \ , \qquad
r_1 =\frac{1}{\ell^2} Li_2(\overline e^\ell\otimes \overline f^\ell)
        \tag 3.5.7
$$
and
$$
R^{(1)} = \left(\sum_{s,t=0}^{4\ell -1}
   \e^{\frac{st}{4}} \omega^s\otimes\omega^t\right) \
 \prod^\ell_{m=0} (1-\e^m \overline e\otimes\overline f)^{-\frac
  m\ell}            \tag 3.5.8
$$

(2) The element $R^{(1)}$ satisfies relations (12),(13).
\endproclaim

{\smc Proof.} First, it ias easy to check that the automorphism
$$
R^{(0)}\left( \left(
\sum^{4\ell -1}_{s,t=0} \e^{\frac{st}{4}}
\omega^s\otimes\omega^t\right) a \left(
\sum^{4\ell -1}_{s,t=0} \e^{\frac{st}{4}}
\omega^s\otimes\omega^t\right)^{-1}\right)         \tag 3.5.9
$$
acts as (1.3.1)--(1.3.3) (assuming there $k^{\pm 1} =
\omega^{\pm 1}e^{\pm\frac z2}$).

Then notice that when $q=e^h\e$, \ $h\to 0$, the asymptotics of the
element (1.2.13) is given by
$$
\exp(\frac{1}{h\ell^2} Li_2(e^\ell\otimes f^\ell))
\prod^{\ell-1}_{m=0} (1-\e^m\overline e\otimes\overline f)^{-\frac m\ell}
(1+O(h))          \tag 3.5.10
$$
This follows from section 3.4.

The theorem now follows from Proposition 1.2.3 and from
(3.2.9),(3.2.10).

\bigskip\bigskip
{\bf 3.6.}Let us discuss the relation of the quasitriangularity
of $\overline{U_\e(sl_2)}$ described in the Theorem 3.5.2 to
the quasitryangularity of the finite dimensional quotient algebra
$$
   U_\e(sl_2)'={\overline{U_\e(sl_2)}}/{<e^\ell,f^\ell,z>}
         \tag 3.6.1
$$
where $\langle e^\ell,f^\ell,z\rangle$
is the ideal generated by these elements.

It is well known that the algebra $U_\e(sl_2)'$ is a quasitriangular finite
dimentional Hopf algebra with the universal R-matrix

$$
    \overline R=\left(\sum_{s,t=0}^{4\ell -1}
   \e^{\frac{st}{4}} \omega^s\otimes\omega^t\right)\
   \sum_{n=0}^{\ell -1}\frac{q^{\frac{k(k-1)}{2}}}{(k)_q!}
    \cdot e^k\otimes f^k          \tag 3.6.1
$$

For details see for example [RT].
The relation between (3.6.1) and (3.5.8) can be described as
follows.

\proclaim{Proposition 3.6.1} (1) The automorphism $R^{(0)}$
>from (3.5.6) induces identity automorphism on the quotient
algebra ${U_\e(sl_2)'}^{\otimes 2}$.
(2) We have the identity:
$$
\prod^\ell_{m=0}(1-\e^m z^m)= \sum_{n=0}^{\ell-1}
\frac{q^{\frac{k(k+1)}{2}}}{(k)_\e!} z^n + O(z^\ell)  \tag 3.6.2
$$
\endproclaim

{\smc Proof.} The first statement follows immediately from
the fact that point $k=1$, $e=0$, $f=0$ is also a symplectic leave
of the Poisson structure (3.3.2) on $SL_2^*$.

The identity (3.6.2) follows from the comparison of the
asymptotics (3.4.1) with the representation (1.4.2) in
case when $q=e^h, \ h\to 0$, and $z^{\ell}=O(h^2)$.

Thus, quasitriangularity of the quotient algebra $U_\e(sl_2)'$
with the universal $R$-matrix (3.6.1)
is inherited from the autoquasitriangularity of
$\overline{U_\e(sl_2)}$ with the universal $R$-matrix (3.5.8).

\bigskip\bigskip\bigskip
\centerline{\bf Conclusion}
\smallskip

We have shown that in an appropriate sense the algebra
$Uq(sl_2)$ is quasitriangular even when $q$ is a root of 1.
This modified notion reproduces known results for appropriate
quotients of $Uq(sl_2)$ [RT],[R].

Notice that functions similar to those which describe $\overline e
\otimes\overline f$ dependence in (3.5.8) had alredy appeared in
the literature in context of Chiral Potts model (see [BB] and [FK]).

The generalization of these results for $Uq(\frak G)$ for simple Lie
algebras $\frak G$ is  straightforward and will be done in a separate
publication.

The following list of problems seems natural to understand now:
\roster
\item"(i)"  The description of the category of modules for
  algebras $U_\e(\frak G)$. One has to understand the category
  where both associative algebra and Poisson structures for
  $U_\e(\frak G)$ are taken into account.
\item"(ii)" The description of algebras $U_\e(\hat\frak G)$ where
  $\hat\frak G$ is a Kac-Moody algebra. It is especially
  interesting to do this for affine Lie algebras $\hat\frak G$.
\endroster
First steps towards understanding question (i) have already been done
in [WX] for $\e =1$.Certain results about quantum affine algebra
$U_q({\hat sl(n)})$ including the description of minimal cyclic
representations can be found in [DJMM].

\bigskip\bigskip\bigskip\bigskip
\centerline{\bf References}

\label{[BB]} Bazhanov V.V.,Baxter V.V. ``New solvable lattice models
  in three dimensions,''{\it Journal of Statistical Physics}
  vol.{\bf 69},no3/4 pp.463-485.
\label {[DJMM]} Date E., Jimbo M., Miki K., Miwa T., ``Generalized
   Chiral Potts Models and Minimal Cyclic Representations of
   $U_q({\hat sl(n)})$,'' {\it Comm Mathem Physics} vol {\bf 137}
   (1991)pp.133-148.
\label{[Dr]} Drinfeld, V. G., ``Quantum groups,'' {\it Proc. Intern.
  Congress of Math.} (Berkeley, 1986), AMS 1987, pp.798--820.
\label{[Dr1]} Drinfeld, V. G., ``On almost cocommutative Hopf
   algebras,'' {\it Algebra and Analysis} vol. {\bf 1}, no. 2,
   pp.30--46.
\label{[DK]}  DeConcini, C. and Kac V., ``Representations of
   quantum groups at roots of 1,'' in {\it Progress in Mathematics}
   vol {\bf 92}, Birkhauser 1990.
\label{[DP]} DeConcini, C. and Procesi, C., ``Quantum groups,''
   preprint no. 6,  Scuola Normale Superiore, 1993.
\label{[FK]} Faddeev, L.D., Kashaev M.,Helsinki preprint,1993.
\label{[FRT]} Faddeev, L. D., Reshetikhin, N. Yu., and
   Takhtajan, L. A., ``Quantization of Lie groups and Lie algebras,''
   {\it Algebra and Analysis} vol. {\bf 1}, no. 1, pp.178--206.
\label{[J]} Jimbo, M., ``$q$-difference analog of $\cap\frak G$ and
   the Yang-Baxter equation,'' {\it Lett. Math. Phys.} vol. {\bf 10}
   (1985), pp.63--69.
\label{[K]} Kac, V. G., ``Infinite dimensional Lie algebras,'',Cambridge
   University Press,1990.
\label{[L]} Lusztig G., {\it Introduction to Quantum Groups}, Progress
  in Mathematics, vol. {\bf 110}, Birkhauser, 1993.
\label{[R]} Reshetikhin N., ``Quasitriangularity of quantum groups
  and Poisson-quasitriangular  Hopf-Poisson algebras,''
  U.C. Berkeley preprint, 1992.
\label{[RT]} Reshetikhin N.,Turaev V., ``Invariants of 3-manifolds
   via link polynomials and quantum groups,'' {\it Invetnt.
   Math.} vol. {\bf 103}(1991), pp.545-597.
\label{[S1]} Sklyanin, E. K., ``On some algebraical structures
   related to the Yang-Baxter equation,'' {\it Funct.Anal.
   and its applications} vol. {\bf 16},no. 4 (1982), pp.27-34.
\label{[S2]} Sklyanin, E. K., ``On one algebra generated by
   quadratic relations,'' {\it Usp. Mat. Nauk.} vol. {\bf 40},
    no. 2 (1985), p.214.
\label{[WX]} Weinstein A. and Xu W., ``Classical solutions to the
  quantum Yang-Baxter equation,'' {\it Comm. Math. Phys.} {\bf 43}
  (1992), 309--344.

\enddocument